# Urban Logistics Dynamics: A User-Centric Approach to Traffic Modeling and Kinetic Parameter Analysis

Emilienne Lardy, Eric Ballot, Mariam Lafkihi

***Abstract*—**Efficient urban logistics requires a comprehensive understanding of traffic dynamics, particularly as it pertains to kinetic parameters influencing energy consumption and trip duration estimations. While real-time traffic information is increasingly accessible, current high-precision forecasting services embedded in route planning often function as opaque 'black boxes' for users. These services, typically relying on AI-processed counting data, fall short in accommodating open design parameters essential for management studies, notably within supply chain management. This work revisits the modeling of traffic conditions in the context of city logistics, emphasizing its significance from the user's point of view, with two focuses. Firstly, the focus is not on the vehicle flow but on the vehicles themselves and the impact of the traffic conditions on their driving behavior. This means opening the range of studied indicators beyond vehicle speed, to describe extensively the kinetic and dynamic aspects of the driving behavior. To achieve this, we leverage the Art. Kinema parameters are designed to characterize driving cycles. Secondly, this study examines how the driving context (i.e., exogenous factors to the traffic flow) determines the mentioned driving behavior. Specifically, we explore how accurately the kinetic behavior of a vehicle can be predicted based on a limited set of exogenous factors, such as time, day, road type, orientation, slope, and weather conditions. To answer this question, statistical analysis was conducted on real-world driving data, which includes high-frequency measurements of vehicle speed. A factor analysis and a generalized linear model have been established to link kinetic parameters with independent categorical contextual variables. The results include an assessment of the adjustment quality and the robustness of the models, as well as an overview of the model's outputs.

***Keywords*—**Factor analysis, generalized linear model, real world driving data, traffic congestion, urban logistics, vehicle kinematics.

## I. Introduction

THE ambition of road traffic models is firstly to describe the vehicle flow through a selection of key performance indicators, such as the throughput and the speed. Secondly, it aims at modeling empirically observed phenomena such as the hysteresis effects or the capacity drop that are generated by traffic jams.

According to the genealogy compiled by [1], traffic flow models have been developed along four branches, which are the fundamental relation, the microscopic models, the mesoscopic models, and the macroscopic models. The common root of these models is the fundamental relations, or fundamental diagrams, between four main parameters which are the vehicle speed, the vehicle spacing, the vehicle flow, and the vehicle density [2], [3]. In particular, the flow as a function of density is initially increasing in the context of free-flowing traffic, until a critical density is reached, at which point it becomes decreasing because of road congestion [4].

The first branch of models is dedicated to refining these fundamental relationships, by specifying their properties, the critical values and thresholds, and other nonlinear phenomena.

The branch of macroscopic models builds on these fundamental relations to describe the vehicles throughput as a continuous flow, drawing inspiration from fluid mechanics. It stems from the Lighthill Whitham Richard (LWR) flow conservation equation, named after the pioneers of this branch [5], [6], which models the flow of vehicles as a homogeneous one-dimensional flow. The simplicity of the LWR model is both its strength, for its robustness and for its ease of calibration, and its weakness because of its inaccuracies in relation to reality. This justifies both to extend the research into more precise variants of this model, and to take an interest in less aggregated models.

The branch of microscopic models traffic flow discretely, by distinguishing individual vehicle speeds and positions, and specifying the relationship with their leaders and followers.

The branch of mesoscopic models is an in-between approach, inspired by gas kinetic models, that describes the flow in terms of probability distributions.

These three types of models, along with hybrid approaches, have been put in practice to study the interactions between urban road traffic and urban freight distribution. In one way, the impact of goods movements on road traffic has been shown to be negative. Indeed, the presence of delivery trucks and vans on the road, with their low speed, their poor maneuverability and their frequent stops for deliveries, leads to a reduction in capacity [7]-[9].

In the other way, the road traffic is also unfavorable to the efficiency and to the costs of delivery tours, since competition for road space and for lane-side parking increases the journey times and the fuel consumption. The most recent literature review to our knowledge shows that more than 75% of the studies since 2001 use microscopic models, amongst which the agent-based simulation models are most popular because of their convenience as decision tools with the available commercial software [10]. The authors conclude the study by suggesting several future research directions, among which, three recommendations will be followed-up in this work.

Firstly, the survey reveals that simulations focus on small scale implementations of innovations, and the assessment of the

E. Lardy, M. Lafkihi, and E. Ballot are with Mines Paris, PSL University, Centre for management science (CGS), i3 UMR CNRS, 75006 Paris, France (e-mail: emilienne.lardy@minesparis.psl.eu).

impacts of supply chain management choices on a wide scale is lacking. Thus, future studies should expand the coverage of simulations. For that purpose, it appears that while the microscopic models currently in vogue have been well suited for their utilization, they are not ideal as a scaling up tool. Instead, this work advocates for user-centric mesoscopic models.

Secondly, the survey points out low transparency on the context data of simulation studies. Yet the context, for example, geographic area or the weekly and hourly period, can lead to significantly different traffic. This goes hand in hand with a lack of studies that include data collection and model calibration stages. There is clearly tension between the need to incorporate and to broaden context factors in the models, and the arduousness of accessing and collecting urban traffic data. This is why this work opens widely the range of context factors under consideration and endeavors to clarify and to hierarchize and their impact on results. The goal is that future studies can target and streamline the data collection process with full knowledge of the situation.

Lastly, the survey exposes the fragmentation of the evaluation of the simulation outcomes. Each study selects a custom set of performance indicators, according to the research requirements, that can cover many facets of freight distribution ranging from customer services to costs and pollutants emissions. This has two consequences: first, the simulation outcomes are biased by the choice of KPI and give a partial rendition of the management choices under review, and second, the heterogeneity of KPI makes it difficult to compare the results between studies. This leads the authors to call for "a comprehensive sets of performance indicators" and "universal guidelines" for future studies. While it seems unlikely that a set-in-stone set of indicators can cover both present and future concerns in supply chain management exhaustively, their common intermediate calculations (speed, driving speed, idle time etc.) could be standardized. As such, this work recommends Art.Kinema parameters, which are a wide range of user-centric kinetic parameters designed to comprehensively describe driving cycles [11].

The Art.Kinema parameters were first created in 2003 as part of the ARTEMIS project (i.e., Assessment and Reliability of Transport Emission Models and Inventory Systems, project funded by the European Commission within the 5th Framework Research Programme, DG TREN) [12]. Because of their adequacy for comparing driving cycles developed all around the world, with different vehicle types and various driving conditions, they are used in a reference book for driving cycles by Barlow et al. in 2009 [11]. Driving cycles have established their high versatility to measure fuel consumption, electric vehicle autonomy and polluting emissions, or to simulate the performances of innovative automotive technologies, and also their longevity as a recognized, high-performance tool [13]. Thus, leveraging the expertise of driving cycles and the Art.Kinema parameters is worthwhile to produce user-centric mesoscopic traffic models.

The objective of this paper is to study user-centric kinematic parameters that describe the driving behavior of vehicles in urban traffic. We open widely the range of context factors that can alter the said urban traffic conditions, considering the time, the day, the road type, the orientation, the slope, and the weather conditions. We explore how accurately the kinetic behavior of vehicles can be predicted based on a limited set of exogenous factors.

This paper is structured as follows; the first section is the introduction. The second section presents the composition of the real-world driving data that is the basis for this study, as well as the analysis methodology. The two following sections present the results, firstly of the exploratory analysis, and secondly of the predictive analysis.

## II. MATERIALS AND METHODS

This work is based on a sample of vehicle traffic recordings under real conditions in Ile-de-France, France. The data were collected from several unidentifiable vehicles between the end of September 2022 and the end of December 2022. It covers a geographical rectangle between Rungis - a major logistics platform - in the south-east corner, and Beaugrenelle - a dense commercial intra-urban area - in the north-west corner, that includes the A6a/b motorway - a major motorway serving Paris. The database consists in 297,201 route segments featuring two types of fields, namely kinetic parameters and context parameters.

The kinetic parameters are Art.Kinema parameters computed over each route segment. To make the analysis more readable, only eight of the 42 variables compiled by Barlow et al. [11] are referenced here: the average speed, the average driving speed, the speed standard error, the average positive acceleration, as well as the percentage of time driving, standing, accelerating, decelerating and cruising. The methods can be extended to the remaining parameters.

The context parameters cover a wide range of external factors including the weather conditions (e.g., dry, rain, etc.), the solar conditions (e.g., dawn, day, etc.), the week period (e.g., week, weekend), the slope intensity (e.g., soft, medium, severe). The geographical coordinates provide information on the type of road, categorized as the six OpenStreetMap standard highway values, plus the Parisian ring road also available as a separate tag. The geo-positioning of measurements also informs on the orientation of the route segment in relation with the urban gradient: it is either incoming to ('in') or outgoing from ('out') the urban center, or non-oriented ('no'). Lastly, the Unix timestamp of each measure informs on the month, the day and the hour of the observations.

Through the prism of traffic modeling, the kinetic parameters are considered to be endogenous variables, i.e., they emanate from traffic conditions, while contextual parameters are exogenous variables, i.e., they are completely determined by external conditions (e.g., time, weather, geography). The goal is to be able to model the kinetic variables as a function of a selection of context variables, in order to be able to predict their value accurately and transparently in a wide range of logistics simulation contexts.

In that aim, the methodology of this work boils down into two steps. An exploratory analysis is first conducted to visualize the dataset and to inform the next step. Context parameters cannot be analyzed one by one, for example by assessing the difference between means with a variance analysis test, because there are many interactions to be expected (e.g., between the hour and the day). Instead, a multidimensional analysis has to be performed, and this is done by means of a factor analysis on mixed data, in accordance with Pagès [14, p.65] and implemented in Python using its built-in linear algebra library Numpy [15] and the Matplotlib library for visualization [16].

A predictive analysis is then carried out to model the expected value of kinetic parameters conditionally on context parameters. In Big Data analytics in transportation systems, linear regression is the most common technique, because it is robust, easy to implement and to interpret [17]. The statistical framework of Generalized Linear Model (GLM) preserves the benefits of linear regression and is more suitable for positive data and positively skewed data, as is the case for kinetic parameters in this work. Thus, it is adopted for this step and it is implemented in accordance with Gills [18] in Python using the statistics library Statsmodels [19].

The dataset is randomly split into a training set, containing 30% of route segments, and a testing set with the remaining 70%. The training set is the basis for the factor analysis in Section III and for the calibration of the GLMs in Section IV, and the testing set is used to assess the GLMs robustness in Subsection IV *B*.

## III. EXPLORATORY ANALYSIS

The results of the factor analysis are displayed on Figs. 1-9, which detail the coordinates of kinetic and context variables on the first two factors. The first two factors explain 37.9% of the inertia of the data points.

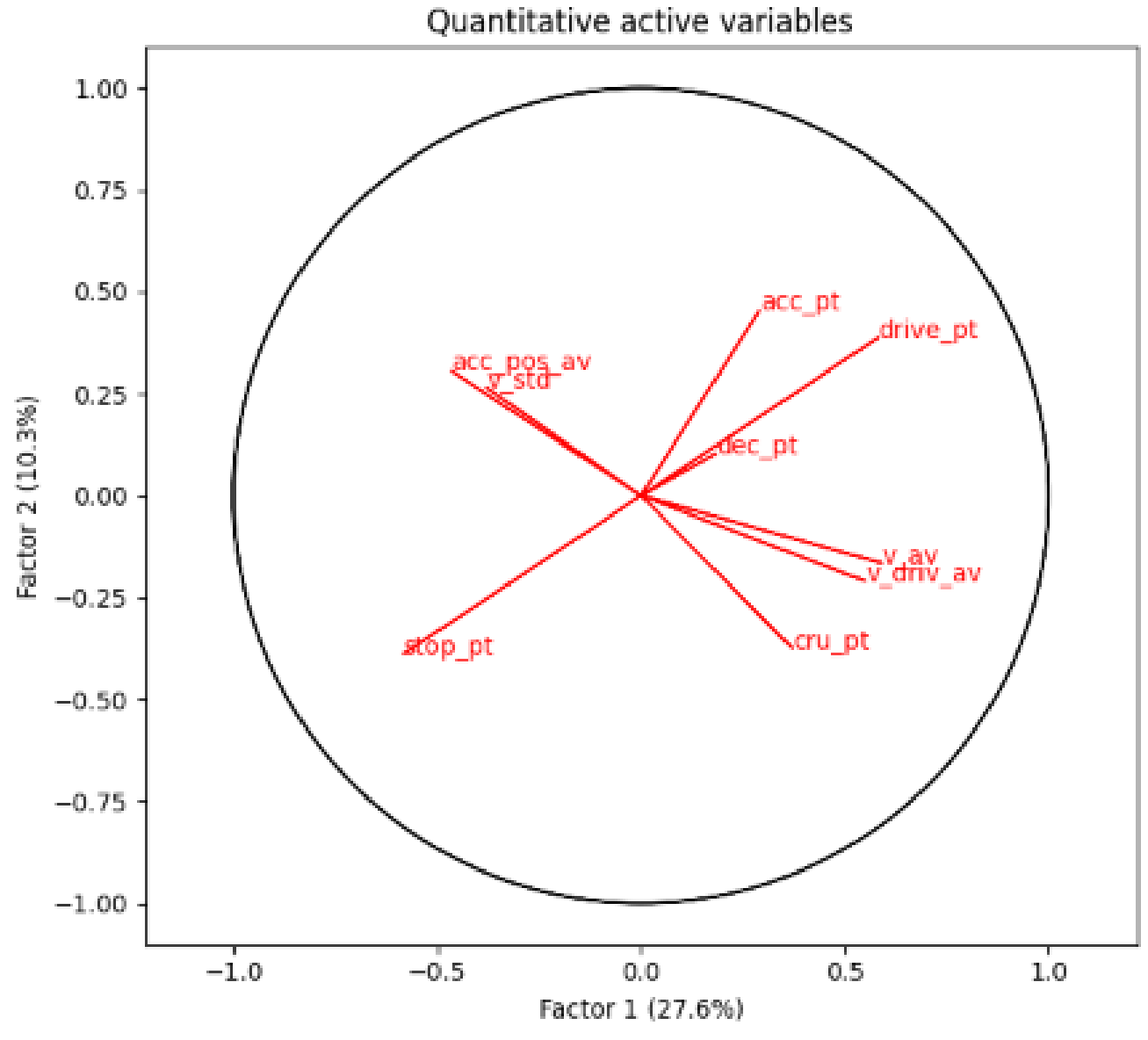

Fig. 1 Correlation circle of the two first factors

The kinetic parameters, which are ordinal, are represented within a correlation circle in Fig. 1. The correlation coefficient of each pair of variables is equal to the cosine of the angle between their representative vectors. This means that narrow angles indicate a high positive correlation, such as for the average speed 'v_av', the average driving speed 'v_driv_av', and the percentage of time spent cruising 'cru_pt'. Angles close to 180° indicate a high negative correlation, as between the average speed and both the average positive acceleration 'acc_pos_av' and the standard deviation of speed 'v_std'. Lastly, angles close to 90° indicate independence, as between the average speed and the percentage of time spent accelerating.

In terms of trip time distribution, it appears that the percentage of time spent driving 'drive_pt', accelerating 'acc_pt', and decelerating 'dec_pt' are positively correlated. They are also negatively correlated with time spent standing 'stop_pt', and relatively independent with time spent cruising 'cru_pt'.

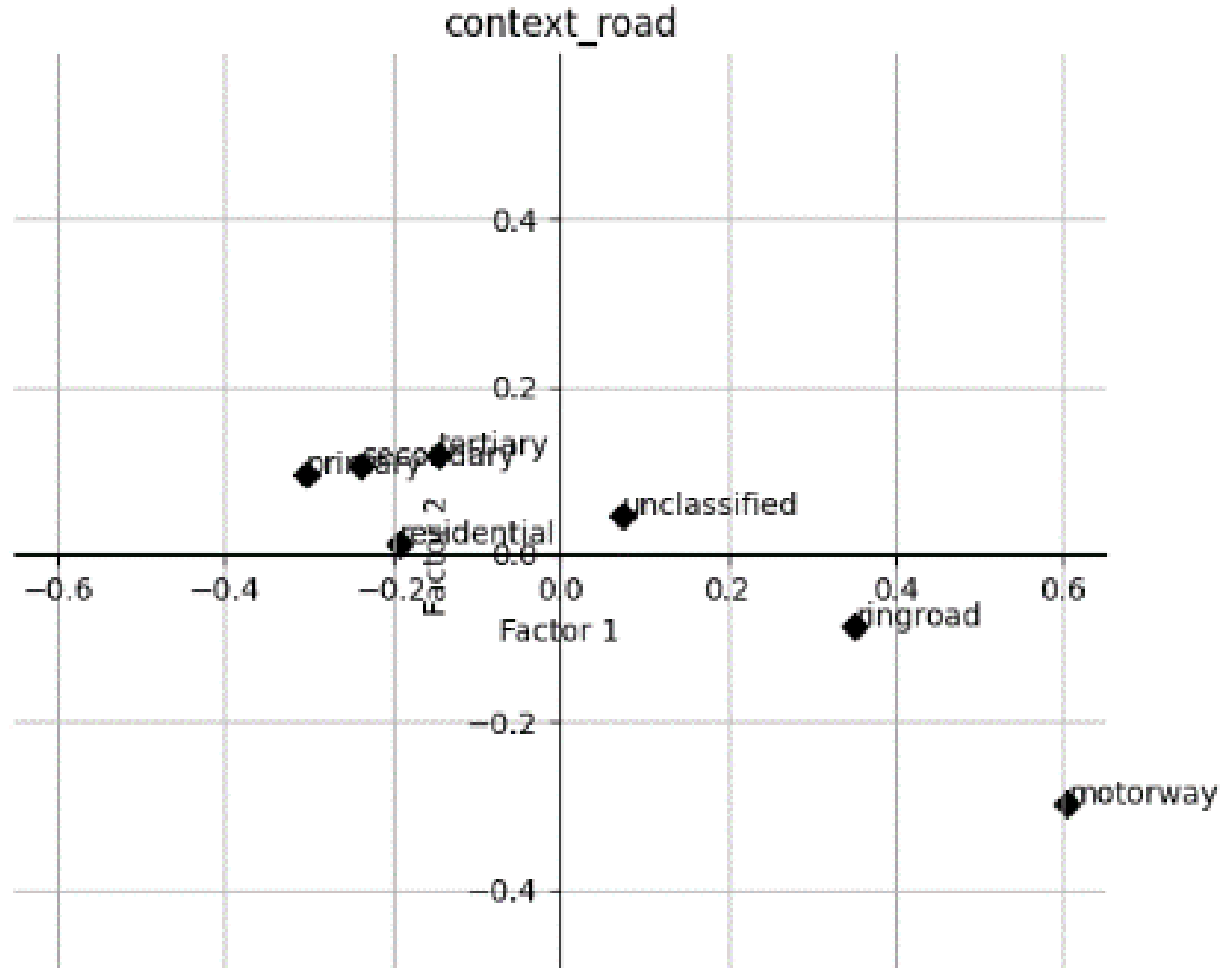

Fig. 2 Coordinates of road categories

The road categories contribute highly to the first two factors, as shown by their high coordinates scattered in Fig. 2. Primary, secondary, tertiary and residential road categories appear to have a similar traffic behavior, as proven by their close coordinates. The ring road and the motorway are off-center, they are associated with higher speeds and cruising durations, since their coordinates are on the bottom right quadrant of the graph just as average speed and percentage of time spent cruising. The 'unclassified' category is difficult to comment because it regroups different situations, some lanes on private properties and some regular roads, which might explain why it has middle grounds coordinates.

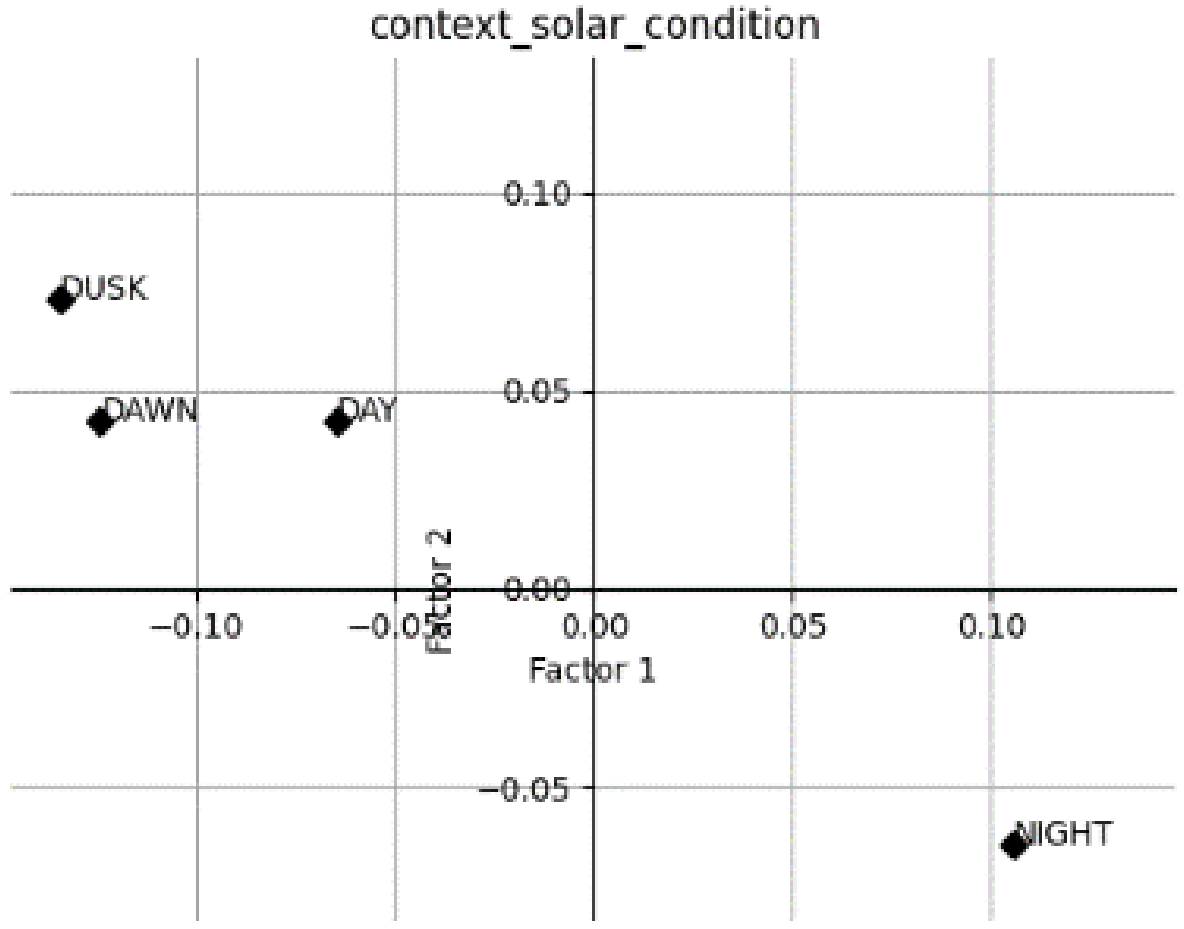

Fig. 3 Coordinates of solar condition categories

In terms of solar conditions, ‘dusk’, ‘dawn’, and ‘day’ have pretty close coordinates, as shown in Fig. 3, while ‘night’ is off-center and associated with higher speeds. It makes sense that the traffic is more free flowing at night, due to a drop in activity and in road usage, rather than to the luminosity. Solar conditions are more relevant to the study of road safety and accidentology [20]; for example, dawn and dusk correspond to periods when the sunrays are close to the horizontal and can hinder the driver's visibility. But they do not naturally correspond to periods of congestion or free-flowing traffic, if anything because they depend on the season, since sunrise and sunset times change from one day to the next, while road usage does not. This is why, in the study of traffic conditions, the hourly profile seems more appropriate, and this is detailed in Fig. 4.

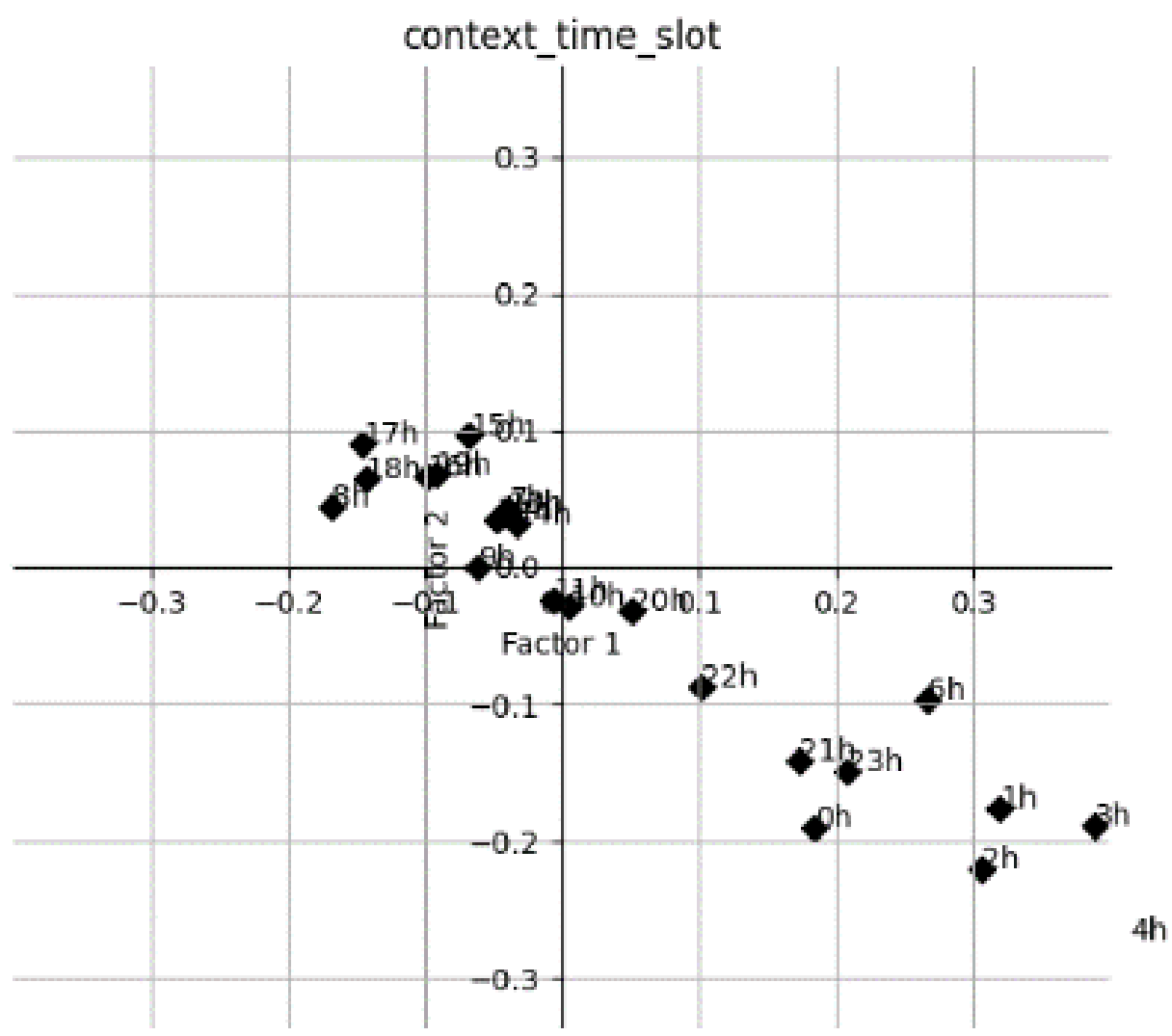

Fig. 4 Coordinates of hourly profile categories

As can be expected from Fig. 3 , Fig. 4 shows that the night hours from 9 PM to 6 AM are associated with higher speeds than other hours. Two peak congestion hours appear, by association with low speeds. First at 8 AM, and to a lesser extent a 7 AM and 9 AM. Second from 3 PM to 7 PM, and to a lesser extent from 12 PM to 2 PM. The slots from 10 AM to 11 AM and at 8 PM are positioned as middle grounds, close to the origin.

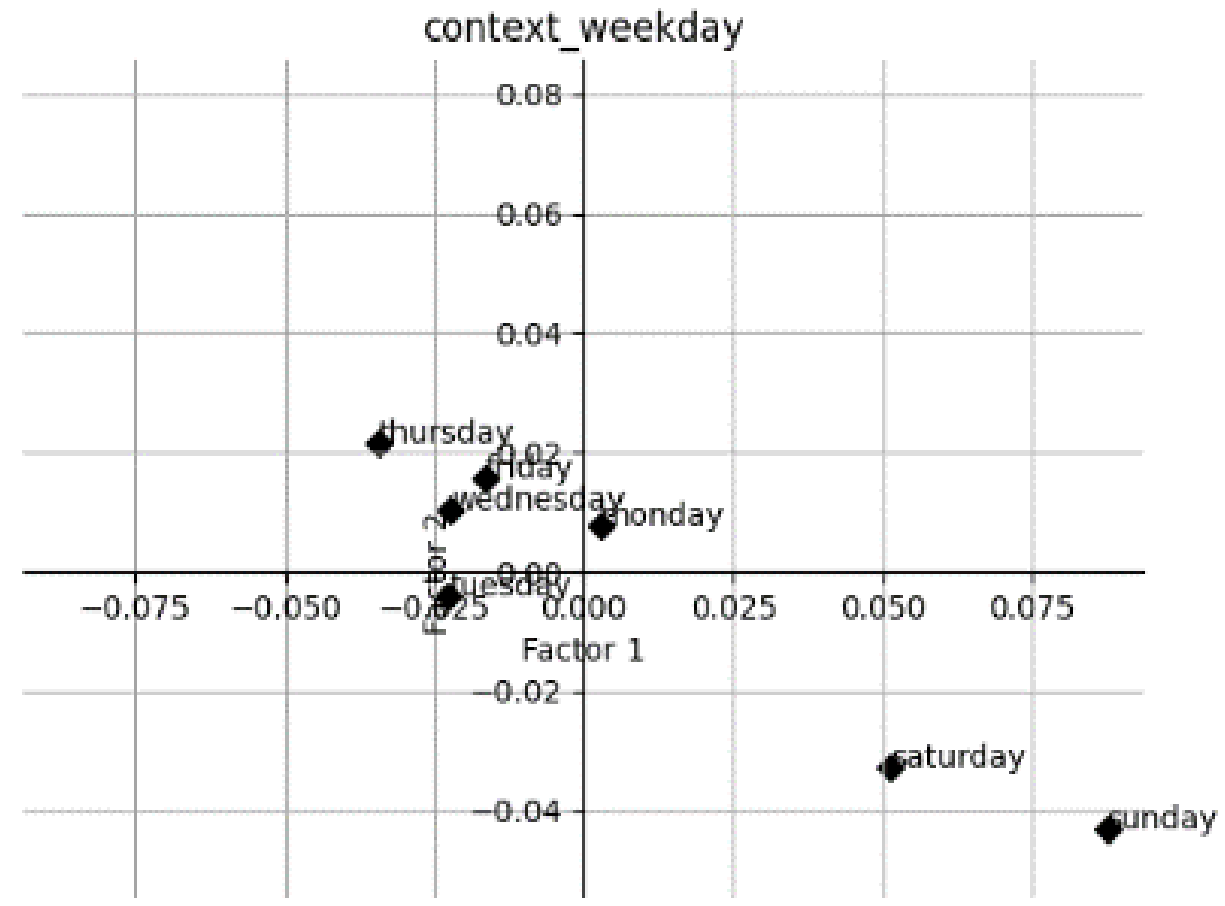

Fig. 5 Coordinates of weekly profile categories

From Fig. 5, the weekly profile categories plainly form two groups: the week day from Monday to Friday on the one hand, and the weekend days Saturday and Sunday on the other. This IS WHY ONLY TWO CATEGORIES ARE ENOUGH TO SYNTHESIZE THE WEEK periods, ‘week’ and ‘weekend’, as shown in Fig. 6.

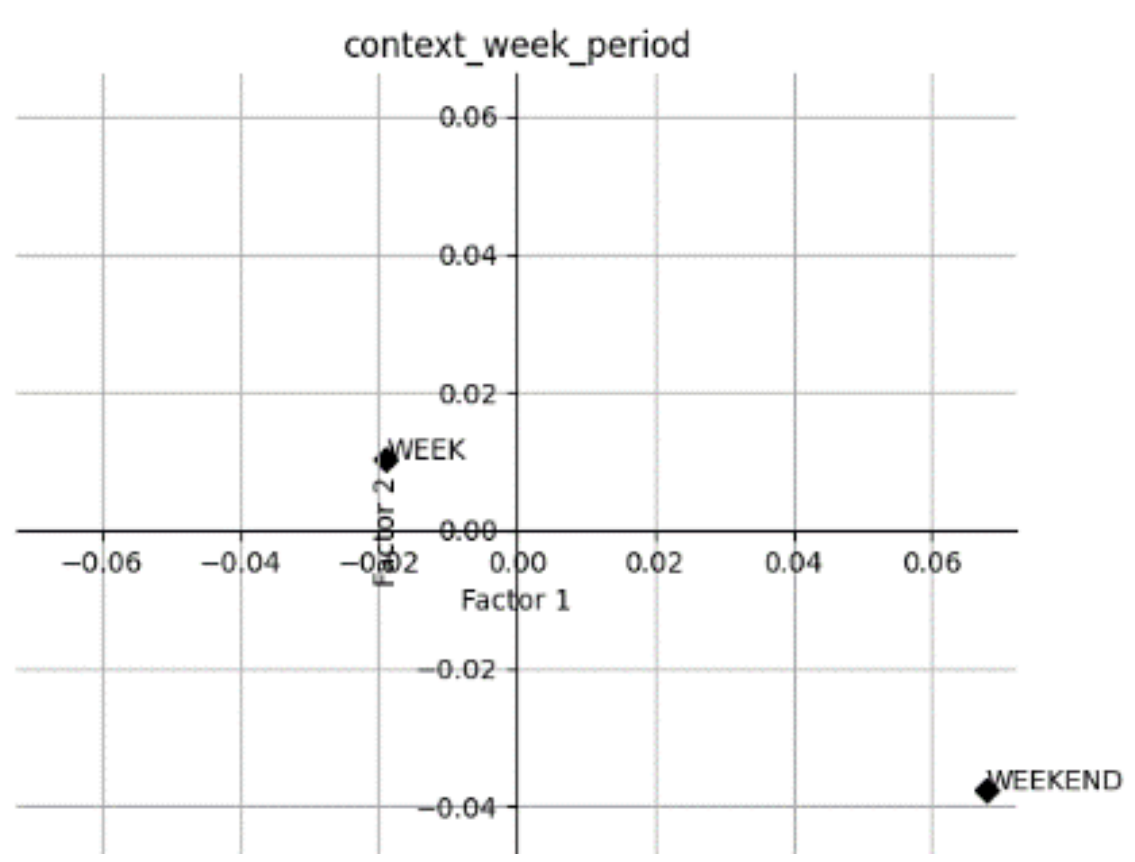

Fig. 6 Coordinates of week periods categories

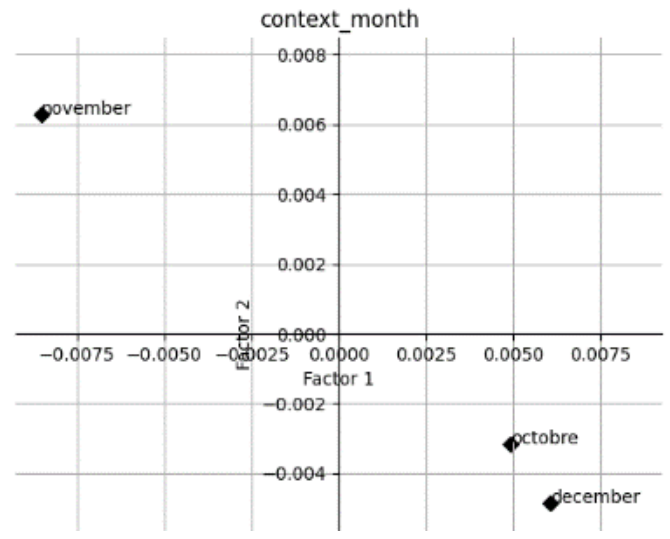

Fig. 7 Coordinates of yearly profile categories

The data collection only covers three months, from October to December 2022, but it seems that the yearly profile has very low contribution to the traffic conditions, as the three categories are positioned extremely close to the origin. Indeed in Fig. *7*, which is zoomed compared to previous graphs, the axes' ticks show that the coordinates are under 0.007 in absolute value on the two first factors. This proves that while hourly and weekly profiles must be included in a user-centric traffic conditions model, the yearly profile can be ignored.

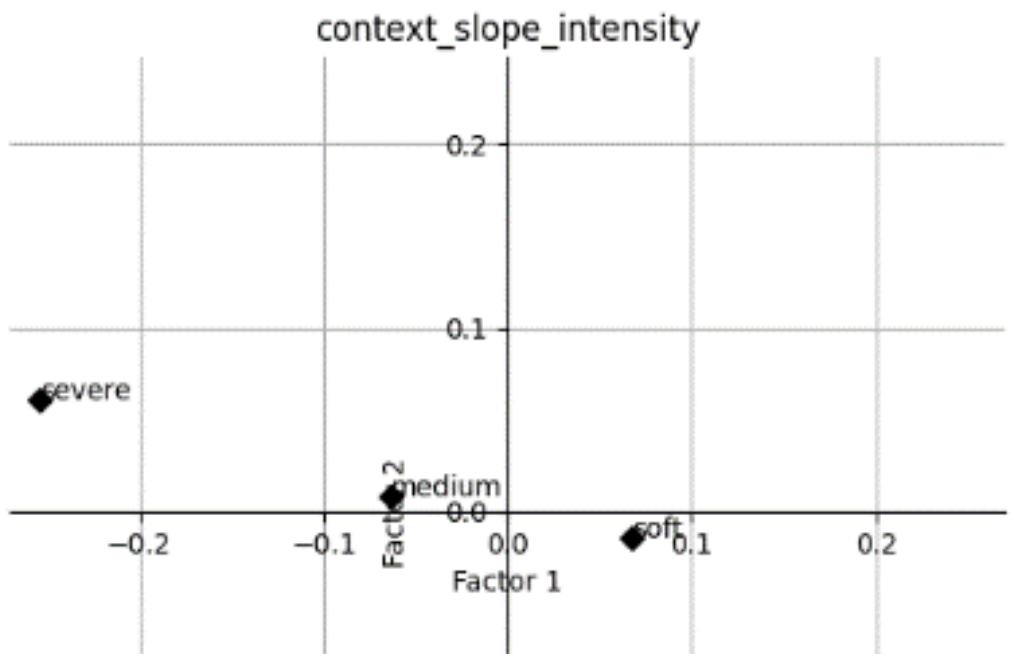

Fig. 8 Coordinates of slope intensity categories

The slope categories scattered in Fig. 8 display the effect of terrain elevation on vehicle speed. Simply put, they show whether the driver manages to maintain a constant speed or if the vehicle tends to accelerate on downhill gradients and decelerates on uphill gradients. The results demonstrate that the effect is significant, and that severe slopes, in the upper left quadrant, are correlated with higher speed standard deviation and high average positive acceleration than medium and soft slopes.

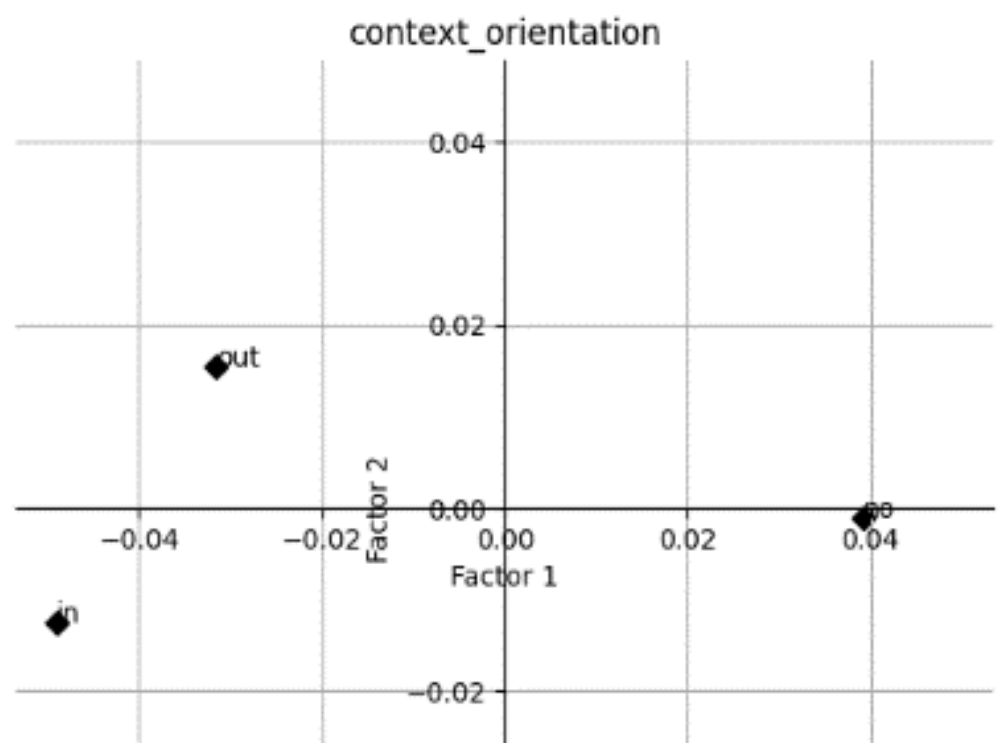

Fig. 9 Coordinates of orientation categories

The 'in' category designates route segments incoming into the urban area. From Fig. 9, it is located in the bottom left quadrant, similar to the percentage of time spent standing ('stop_pt) in Fig. 1. This covariance could be explained by frequent traffic jams while entering the city of Paris. The 'out' category is in the upper left quadrant, along with the standard deviation of speed and the average positive acceleration parameters. This might indicate a heavy traffic, not congested enough to bring vehicles to full stops but still promoting erratic vehicle behaviors.

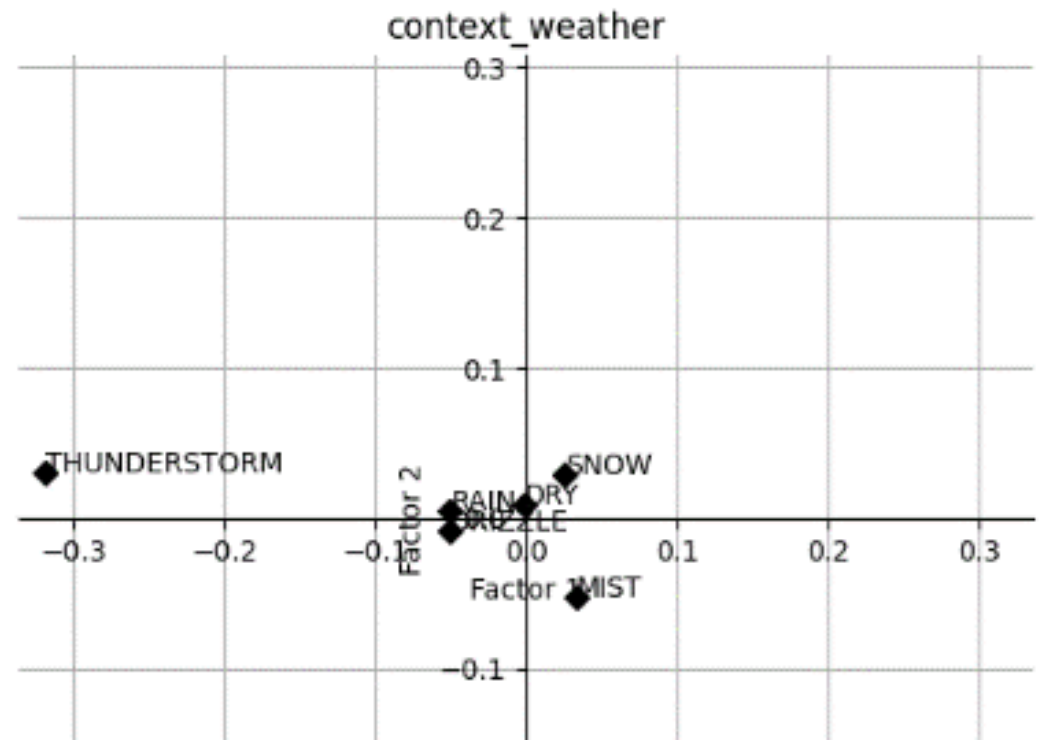

Fig. 10 Coordinates of weather categories

The weather has a low but non negligible impact on traffic conditions, as most categories are gathered around the origin of the graph on Fig. 10, with moderately low coordinates. The 'thunderstorm' category clearly stands out from the other categories, but it is a rare value within the data points, and its off centered coordinates could be an effect of over-representation, which is a risk with rare values in factor analysis. For the same reason, the coordinates of the 'snow' category must be treated with caution. The 'rain' and 'drizzle' categories being close seems reasonable, but interestingly the 'mist' modality is not in the same quadrant, meaning it leads to different traffic behaviors.

## IV. Predictive Analysis

### *A. Implementation*

The implementation choices for the GLM directly stems from the exploratory analysis. Six endogenous variables are selected to compose the linear predictor, namely road, time slot, week period, slope intensity, orientation, and weather categories. Categories with close coordinates are regrouped to prioritize the efficacy of the calibration and of the analysis, and to avoid any over-adjustment risk. This is detailed in Table I.

The linear predictor is a second-order linear combination of exogenous variables in order to investigate the interactions between them. Additionally, the endogenous variables are positive and positively skewed, which calls for a Gamma underlying distribution with a logarithmic link function in the GLM.

TABLE I
CATEGORIES IN THE LINEAR PREDICTOR FOR GLM

| Exogenous variable | categories |
|---|---|
| context_road | • motorway<br>• ringroad<br>• downtown<br>{primary, secondary, tertiary, residential}<br>• unclassified |
| context_time_slot | • peak_hour<br>{8h, 15h, 16h, 17h, 18h, 19h}<br>• near_peak<br>{7h, 9h, 12h, 13h, 14h}<br>• off_peak<br>{10h, 11h, 20h, 21h, 22h}<br>• nighttime<br>{23h, 0h, 1h, 2h, 3h, 4h, 5h, 6h} |

| context_week_period | • WEEK<br>• WEEKEND |
|---|---|
| context_slope_intensity | • soft<br>• medium<br>• severe |
| context_orientation | • in<br>• out<br>• no |
| context_weather | • DRY<br>• MIST<br>• ADVERSE {RAIN, DRIZZLE, SNOW, THUNDERSTORM} |

### B. Adjustment quality

The GLM results are detailed in Table II for four kinetic parameters: the average speed ‘v_av’, the average driving speed ‘v_driv_av’, the standard deviation of speed ‘v_sd’, and the average positive acceleration ‘acc_pos_av’.

The deviance is the goodness of fit indicator for statistic models fitted with maximum likelihood techniques. The likelihood ratio test assesses the goodness of fit of the model by comparing the deviance with the null deviance (i.e., model with only an intercept parameter, the mean of the data). For a GLM, the statistics used is the difference in deviances divided by the scale and by the number of extra terms in the linear predictor [21]. It follows an F statistic, and a p value under 0.05 indicated that the model provides substantial information compared to the overall mean. Here, this is the case for all four variables.

In order to assess the robustness of the model, we calculate the slope between the endogenous variables and the predictions, for both the training set and the testing set. These two slopes, named ‘Slope train’ and ‘Slope test’ in Table II, are equal for the four endogenous variables, which shows that the model is very robust. However, it also uncovers a very important difference between speed intensity variables (i.e., ‘v_av’ and ‘v_driv_av’) and speed dispersion variables (i.e., ‘v_std’ and ‘acc_pos_av’). The slopes values are 0.42 for the former, which indicates is an excellent fit, but they are only 0.06 and 0.13 for the latter, which indicates very high residual variability. In short, while the models are significant and robust for all four variables, they are much more precise in the prediction of speed intensity variables than speed dispersion variables.

TABLE II
GLM REGRESSION RESULTS

| Variable | v_av | v_driv _av | v_std | acc_pos _av |
|---|---|---|---|---|
| Model | GLM | GLM | GLM | GLM |
| Family | Gamma | Gamma | Gamma | Gamma |
| Link | Log | Log | Log | Log |
| Method | IRLS | IRLS | IRLS | IRLS |
| Degrees of Freedom | 89077 | 89077 | 89077 | 89077 |
| Deviance | 49428 | 31705 | 43673 | 30217 |
| AIC | 547728 | 534110 | 270048 | 102017 |
| BIC | -965887 | -983611 | -971642 | -985098 |
| Likelihood Ratio Test | 514.1 | 565.4 | 79 | 182.7 |
| LRT p value | 0 | 0 | 0 | 0 |
| Slope train | 0.42 | 0.42 | 0.06 | 0.13 |
| Slope test | 0.42 | 0.42 | 0.06 | 0.13 |

### C. Prediction results

The predictions result for the same four kinetic parameters are displayed from Fig. 11 to Fig. 14. To avoid cluttering the chart, it would be impractical to plot all predictor combinations, hence, outcomes are presented as a deviation from a reference state. The reference state is as following: the road context is ‘downtown’, the weather context is ‘DRY’, the week period context is ‘WEEK’, the time slot context is ‘off-peak’, the slope intensity context is ‘soft’, and the orientation context is ‘no’. The marginal effect of each category is dispatched along radar plots.

The results include the model prediction, that is to say the expected value of the endogenous variable conditionally on the exogenous variables, as well as a 90% confidence interval for the prediction. These three envelopes are superimposed on the radar plots and labeled as ‘Prediction’ for the expected value, ‘CI upper (95%)’ for the upper value (i.e., the outcome is lower than this upper bound with a 95% confidence level), and ‘CI lower (95%)’ for the lower value (i.e., the outcome is greater than this lower bound with a 95% confidence level).

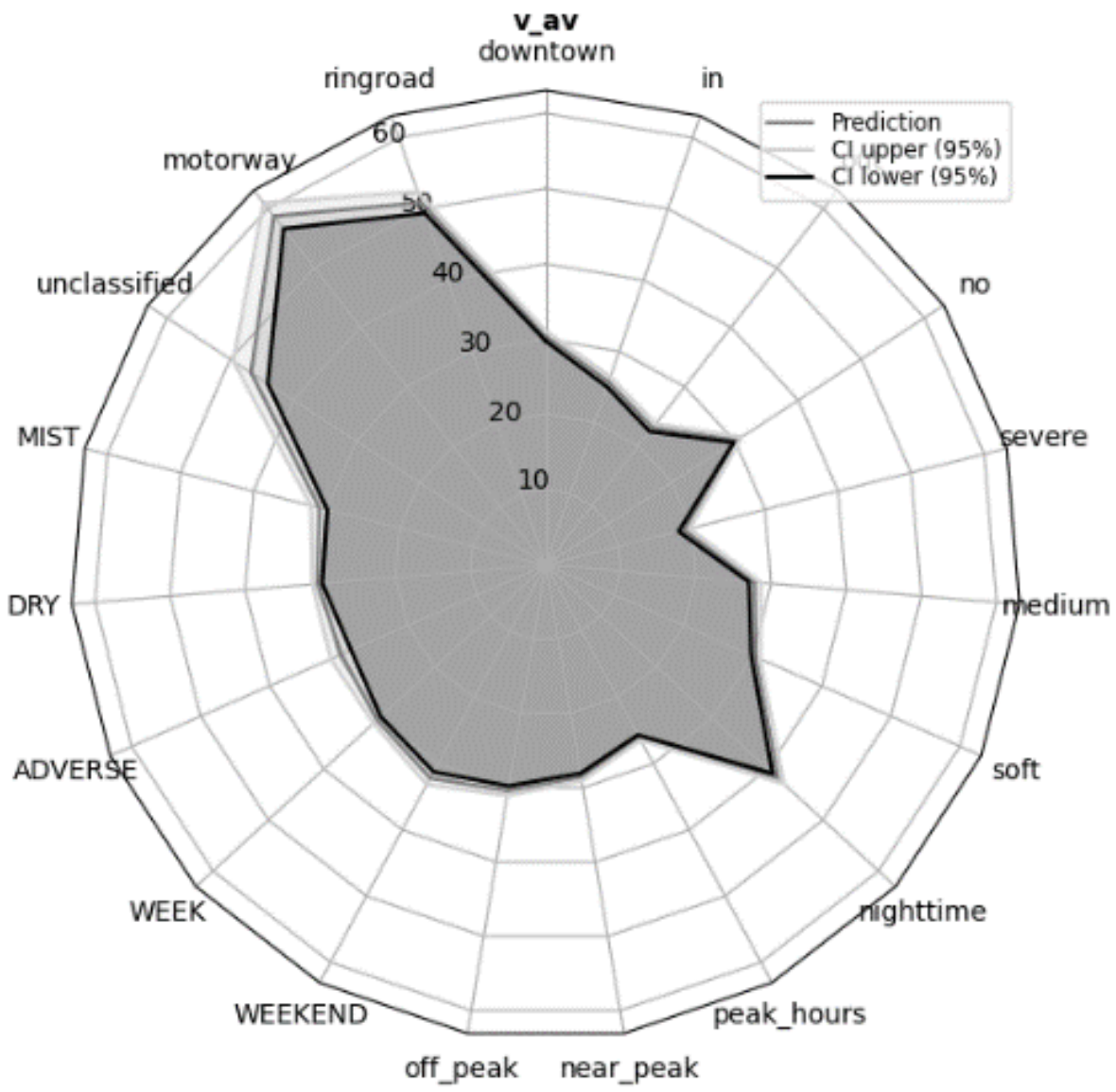

Fig. 11 Prediction results for the average speed in km/h

From Fig. 11, it appears that the most impactful categories are the road types, as the expected average speed is 58 km/h on the motorway and 50 km/h on the ring road, and only 30 km/h downtown. The nighttime presents the opportunity to increase average speed by 10 km/h in the reference state, while peak-hours are responsible for a decrease of around 5 km/h. The weather conditions and the week period have a minimal impact on average speed.

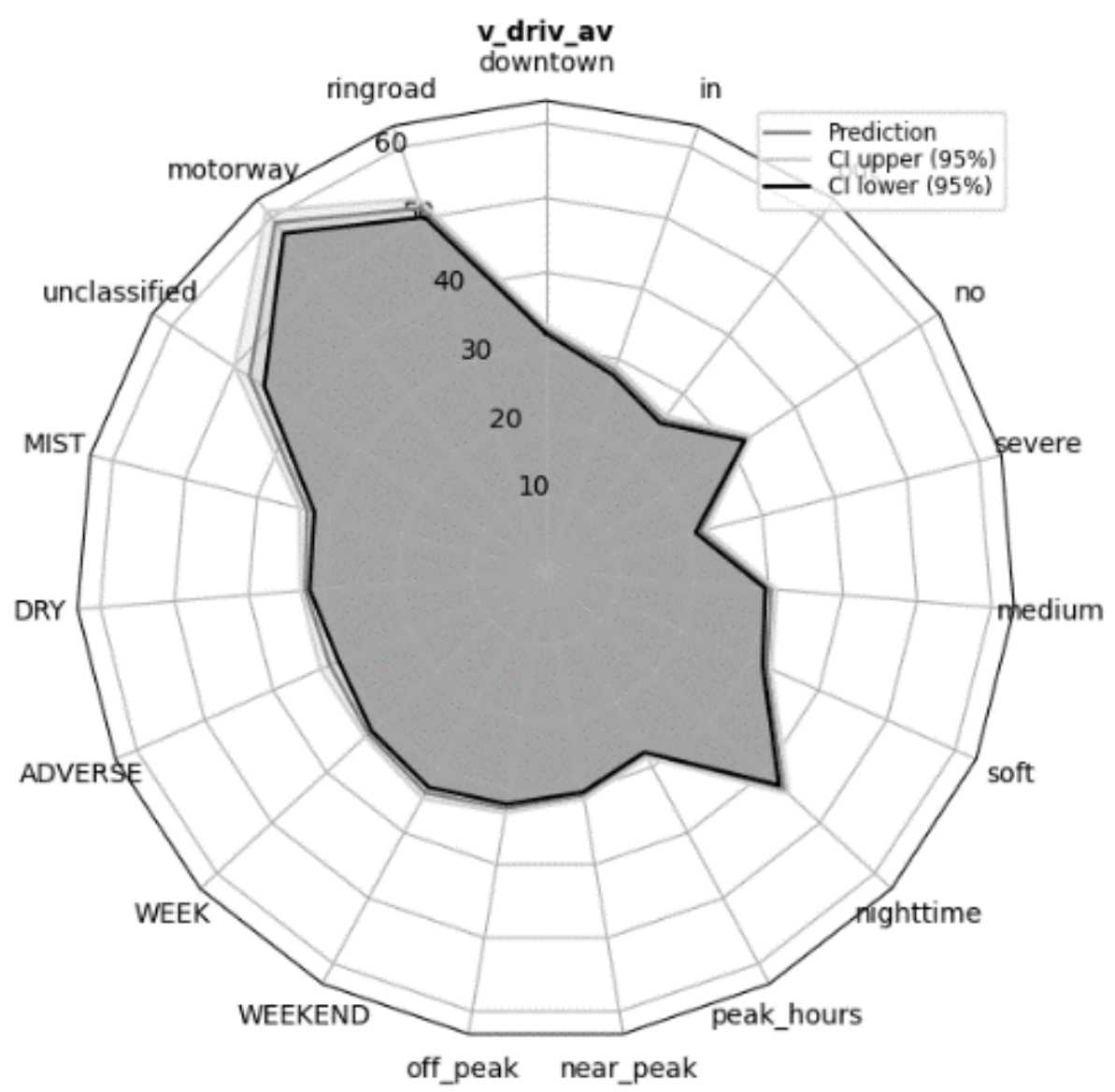

Fig. 12 Prediction results for the average driving speed in km/h

The average driving speed follows the same tendencies as the average speed according to Fig. 12. While it is not very noticeable from their respective graphs, the average driving speed is always slightly higher than the average speed, of about 10%.

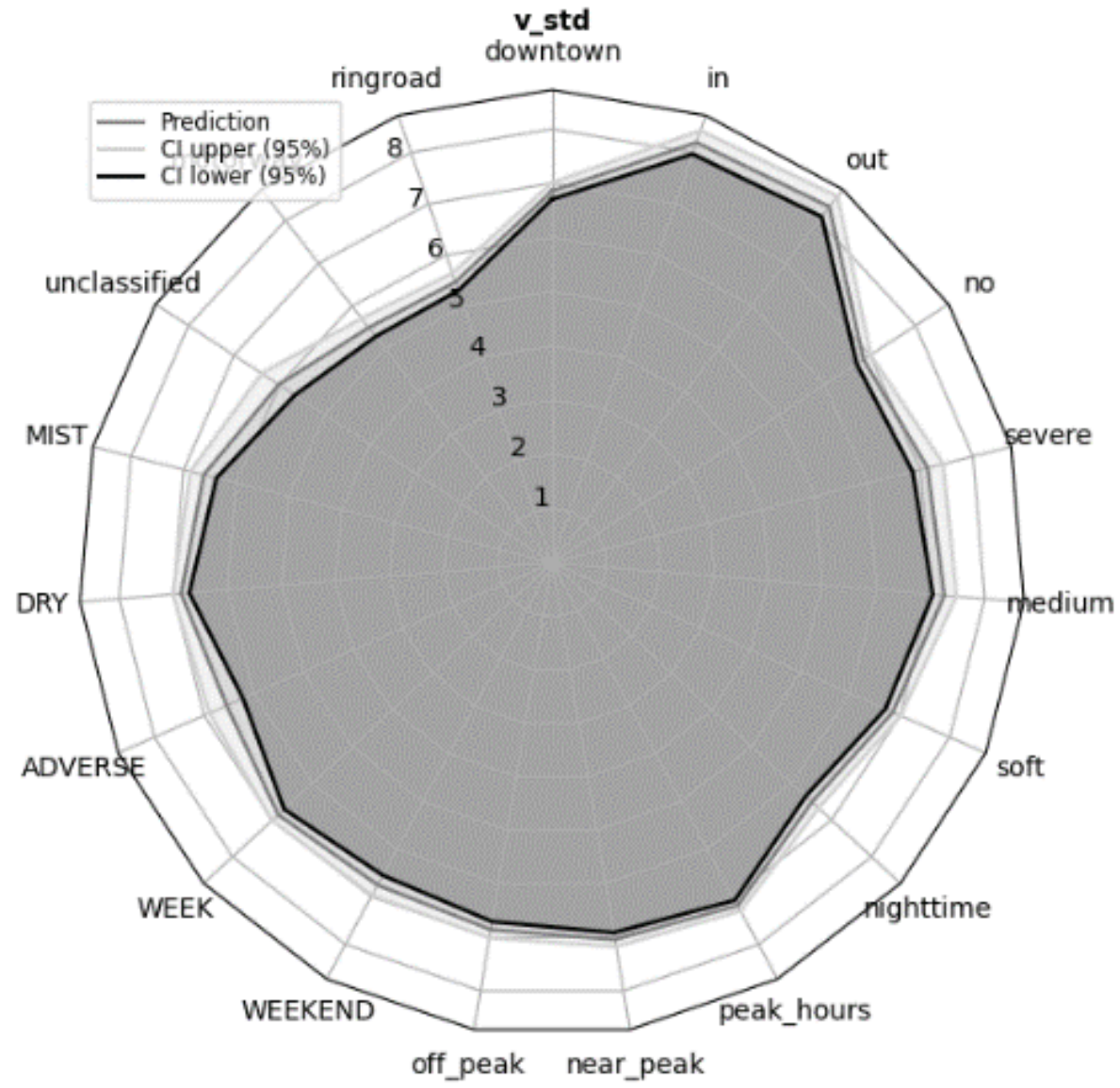

Fig. 13 Prediction results for the standard deviation of speed

The standard deviation of speed does not vary as much as the average speed itself under different context factors, as can be seen from Fig. 13. But the ‘in’ and ‘out’ orientations, as well as the ‘severe’ and ‘medium’ slope intensity, appear to increase the speed dispersion. On the contrary, ‘motorway’ and ‘ring road’ road types, as well as the ‘nighttime’ time slot, are prone to produce more constant speeds.

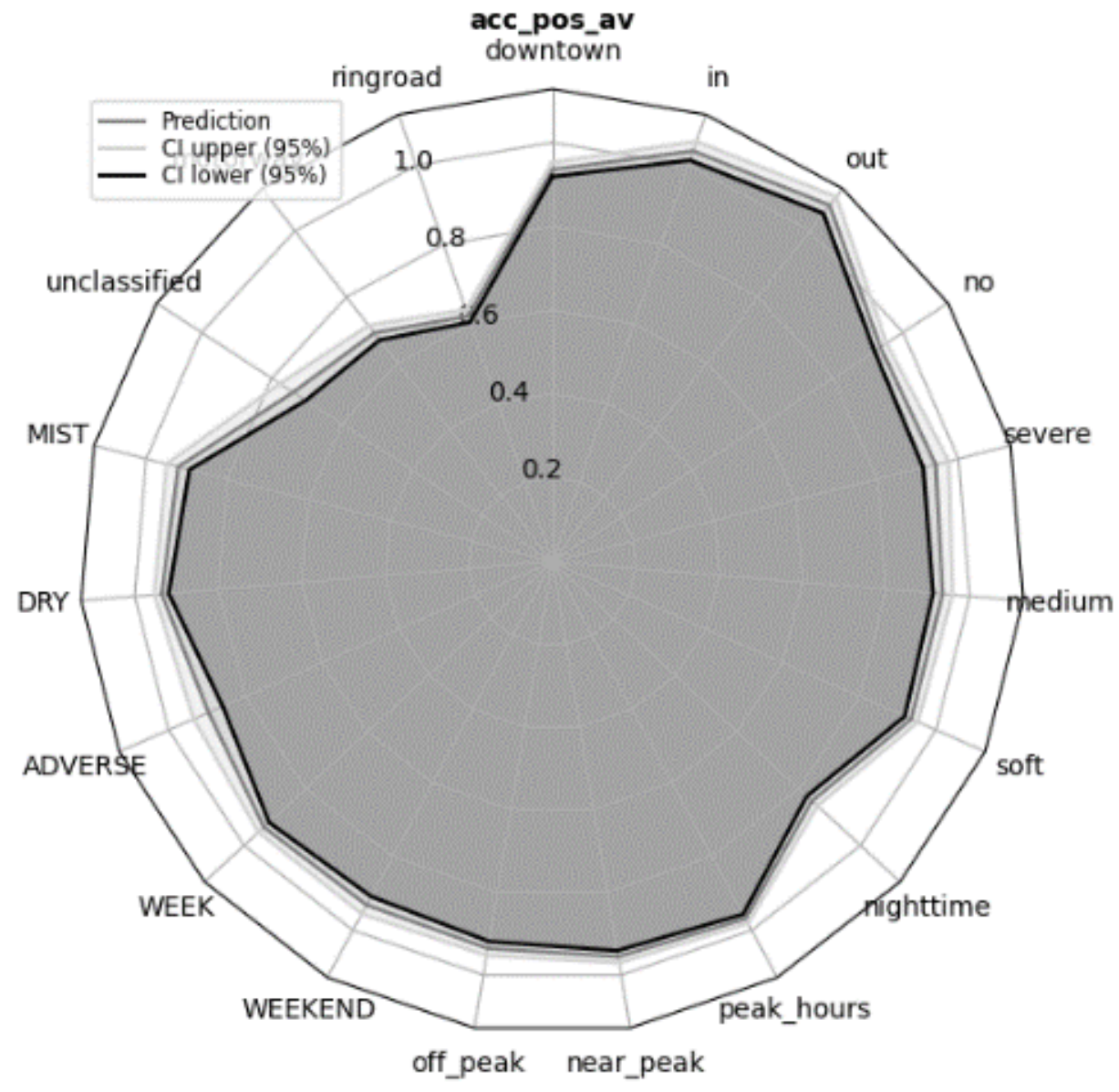

Fig. 14 Prediction results for the average positive acceleration (m.s$^{-2}$)

Lastly, Fig. 14 proves that the average positive acceleration follows similar tendencies as the standard deviation of speed. The expected values range from 0.6 m.s$^{-2}$ to 1.2 m.s$^{-2}$ which is very coherent with current vehicle acceleration capacities.

## V. Conclusion

This work advocates to leverage the Art.Kinema parameters, primarily descriptive of driving cycles, in order to produce a user-centric mesoscopic traffic model. This has major benefits to further the research on the impact of traffic conditions on urban supply chain design and management. Firstly, as a mesoscopic model it allows larger scale simulation than microscopic models. Secondly, their high versatility makes them ideal candidates to somewhat standardize intermediate simulation outcomes, in order to favor comparison between studies.

This work attempts to model these kinematic parameters as a function of a wide range of external factors, in order to target the most impactful ones. The road type, the time of the day, the day of the week, the slope intensity, the orientation relatively to the urban gradient, and the weather all have an established impact on the traffic behavior. A GLM can provide a significant and robust prediction for these kinematic parameters conditionally on external factors. Based on the first results, the quality of the prediction varies from one parameter from another. While the standard error of speed and the average positive acceleration are not fully explained, the average speed and the average driving speed show an excellent observation versus prediction fit.

This work is mainly limited by the lack of similar driving data from other cities and other countries to prove to what extent the results can be extended to urban areas of different sizes and locations.

The results echo some existing supply chain management strategies, for example the night deliveries that take advantage of higher driving speeds at night to shorten the duration of delivery tours. In that vein, the results open up new prospects to increase efficiency and costs of urban freight movements, and further works are needed to put them in practice.

## Acknowledgment

The authors thank the Physical Internet Chair and its funding partners, including GEODIS, as well as the collaborating partner that provided the field data required for this research project.